# Interfacial Reconstructions and Engineering in III-V@II-VI Core-Shell Quantum Dots

*Jordi Llusar,[1] * Abdessamad El-Adel,[1] Luca De Trizio,[2] Liberato Manna,[3] Zeger Hens,[4,5] and Ivan Infante[1,6]**

[1]BCMaterials, Basque Center for Materials, Applications, and Nanostructures, UPV/EHU Science Park, Leioa 48940, Spain.

[2]Chemistry Facility, Istituto Italiano di Tecnologia, Via Morego 30, 16163 Genova, Italy

[3]Nanochemistry, Istituto Italiano di Tecnologia, Via Morego 30, 16163 Genova, Italy

[4]Physics and Chemistry of Nanostructures, Ghent University, B-9000 Gent, Belgium

[5]Center for Nano- and Biophotonics, Ghent University, B-9000 Gent, Belgium

[6]Ikerbasque Basque Foundation for Science, Bilbao 48009, Spain.

## ABSTRACT

In core@shell quantum dots (QDs), the interface between semiconductors of different chemical character largely determines their optoelectronic properties. In III–V@II–VI systems, this boundary involves pronounced chemical and electronic discontinuities that can generate trap states even under complete surface passivation. Using density functional theory on atomistic models of InAs@CdSe QDs, we systematically reconstruct atomic arrangements at the surface and interface to evaluate how local coordination and interfacial dipoles influence the electronic structure. Abrupt interfaces induce charge imbalance and band-gap collapse, whereas introducing an alloyed interlayer that mixes core and shell atoms and vacancies restores energetic alignment and yields delocalized band-edge states, consistent with experimental findings. We also introduce a charge-flow analysis that quantifies charge redistribution across the QD, providing a framework for realistic modeling of interlayer formation and predictive design of defect-free interfaces in core@shell architectures.

Colloidal quantum dots (QDs) are size-tunable semiconductor nanocrystals with discrete energy levels, whose strong carrier confinement yields size-dependent absorption and emission across the visible and infrared (IR) spectrum[1]. However, their large surface-to-volume ratio makes them particularly susceptible to the presence of surface-localized states[2–4], which strongly influence their electronic properties. Surface states can act as traps for electrons or holes and induce non-radiative electron-hole recombination, thereby shortening charge-carrier lifetimes, and reducing the photoluminescence quantum yield (PLQY)[4–6]. In core-only QDs, these states are commonly associated with surface atoms that are dangling, i.e. undercoordinated, introducing localized states within the bandgap[7,8]. Although it has been demonstrated that not all undercoordination leads to trap states[9], reduced coordination generally correlates with higher trap propensity; thus, the coordination number serves as a simple, practical and necessary – though not sufficient – descriptor.

A common strategy to mitigate surface traps is ligand passivation, using ligands with L-type, such as amines, phosphines, or thiols, or X-type functionalities, for instance carboxylates, phosphonates, thiolates or halide, to bind and saturate dangling bonds[10–12]. While ligands stabilize QDs and suppress non-radiative recombination, they are chemically diverse, labile, and sensitive to the environment, complicating long-term control[5,11]. A more robust approach is the growth of an inorganic shell, which embeds inorganic QD core atoms into a fully coordinated inorganic environment[13,14]. While this suppresses core-localized traps, it simultaneously creates a new outer surface that remains susceptible to dangling bonds or structural defects[7]. In type-I heterostructures, core confinement reduces the overlap of carriers with these surface states when increasing the shell thickness[15], allowing a substantial improvement of the photoluminescence efficiency. On the other hand, for borderline type-I cases or for type-II heterostructures, one of the carriers may leak or reside in the shell, in which case surface traps remain problematic [16–19].

Beyond these surface-related considerations, the growth of a shell also introduces a heterointerface. Even in fully coordinated environments, the boundary between chemically distinct core and shell materials may induce local dipoles arising from differences in electronegativity, bonding character, or oxidation state of the constituent ions. These dipoles distort local electrostatic potentials and may alter the alignment of molecular orbitals at the interface, potentially affecting carrier localization together with recombination dynamics and likely inducing the formation of trap states, even if the core is fully coordinated[3,20,21]. Unlike surface traps, which can be probed experimentally by surface-sensitive techniques, such as scanning tunneling microscopy or X-ray photoelectron spectroscopy[3], interfacial traps are buried within the shell and, despite recent advances in solid-state NMR[22,23], remain experimentally difficult to access[21]. This uncertainty motivates the use of theoretical modeling to elucidate the processes occurring at the heterointerface.

Core@shell QDs of the same material family (e.g., II-VI@II-VI) often suppress surface traps and improve stability by encapsulating the core preferably with an epitaxial, lattice-matched shell. The presence of isoelectronic ions in both core and shell makes the transition between the two materials less problematic. Conversely, III-V@II-VI heterostructures amplify interfacial effects due to larger chemical, electronic and structural differences at the core-shell boundary. To investigate these, we decided to analyze InAs@CdSe QDs. Among possible III-V core materials, InAs stands out as a prototypical choice due to its extensive use in IR optoelectronics. At the same time, InAs surfaces are highly prone to trap formation as dictated by the negligible PLQY when not shelled, making them a stringent test case for evaluating passivation strategies[24–26]. Pairing InAs with a II–VI shell such as CdSe is also particularly appealing from the theoretical standpoint: their near-zero lattice mismatch[27] (zinc blende; $a_{InAs}$ = 6.058Å; $a_{CdSe}$ = 6.077Å) suppresses strain effects, making the occurrence of interfacial traps mostly related to their chemical and electronic dissimilarity[28]. This combination makes InAs@CdSe an ideal model system to disentangle the relative contributions of atomic coordination and interfacial dipoles to trap formation.

Building on this, we present a density functional theory (DFT)-based framework to analyze these effects in InAs@CdSe QD models comprising over 2000 atoms (~4.3 nm in overall size). Our approach captures both the magnitude and direction of dipoles at the heterointerface and reveals how different local atomic environments can strongly influence the presence of electronic states within the bandgap. By systematically exploring vacancy and atomic patterns at the interface, we show that it is possible to construct QD models that suppress trap states and provide delocalized, bulk-like wavefunctions. These features are indicative of a more realistic and chemically consistent electronic structure, in alignment with experimental findings[29–31]. These QD models indeed point to the presence of a mixed In-Zn-Se interlayer, one monolayer (ML) thick, as demonstrated by Zhu and coworkers at the InAs@ZnSe core-shell interfaces[19,32]. Based on the results proposed here, we extend the idea to InAs@CdSe and by systematically manipulating interfacial patterns we provide a practical route to more realistic heterostructured QD models. Our vision is that these models will guide future experiments aimed at tailoring interfaces and improving optoelectronic performance.

DFT calculations were carried out with the CP2K quantum chemical software package using the HLE17 exchange-correlation functional due to its good lattice parameter and bandgap estimation[33] (see **Section S1** in the **Supporting Information** – **SI**). Charge distribution across the QD models was quantified from single ground-state calculations: (i) extracting valence Bader charges[34–36] to obtain per-atom deviations from the neutral valence $Z$; (ii) constructing a bond graph from interatomic distances using Cordero et al.'s single-bond covalent radii[37]; and (iii) computing the charge difference between ion pairs by solving a regularized least-squares system – an overview of the workflow is shown in **Scheme S1** in the **SI** . This procedure enforces global neutrality of the isolated QD system (fulfilling charge-balance criterion), such that charge-gains/-losses at an atomic site are balanced by inflow or outflow of charge along its chemical bonds with nearby atoms. Summing such per-bond charge differences along all bonds of an atom defines a charge-flow vector on each atom: its direction indicates whether the atom accepts or donates charge; the orientation indicates where the charge moves from/to, and its magnitude quantifies the net charge transferred from/to it. Full methodological details and the formal connection to classical electric dipoles are provided in **Section S2** in the **SI**.

We first explored an InAs core-only QD, which serves as the substrate for the subsequent CdSe shell growth. Using the nanocrystal builder[38], the model was constructed as a truncated tetrahedron with cation-rich ⟨111⟩, anion-rich $\langle\overline{111}\rangle$, and ⟨100⟩ facets typically observed in TEM images[39–42]. The resulting pristine InAs QD model has a height of ~2.1 nm and is charge-balanced with Cl ligands[39], see **Figure 1a – left**. This type of model presents an optical gap of 1.29 eV defined by the frontier delocalized valence band (VB) and conduction band (CB) edges. The region energetically above the delocalized VB edge however exhibits an accumulation of trap states due to the presence of 3-coordinated (3c) As atoms at the $\langle\overline{111}\rangle$ facets, which lead to facet-specific localization as evidenced in **Figure 1b,c – left**. As demonstrated in reference[32], these facet-specific traps indeed emerge from the accumulation of negative charge on these 3c-As ions. To remove these traps from the model, we reconstructed $\langle\overline{111}\rangle$ facets by exchanging two thirds of the 3c-As atoms by 3c-Cl atoms, see **Figure 1a – right**. This reconstruction reduces the number of 3c-As atoms and, more importantly, diminishes the accumulation of negative charge, since a -3 ion is now replaced by a -1 ion. This leads to the elimination of the facet traps and to full delocalization of the band-edge orbitals (**Figure 1b,c – right**). This improvement is apparently driven by site-specific reconstruction rather than enhanced passivation (**Figure S3e**).

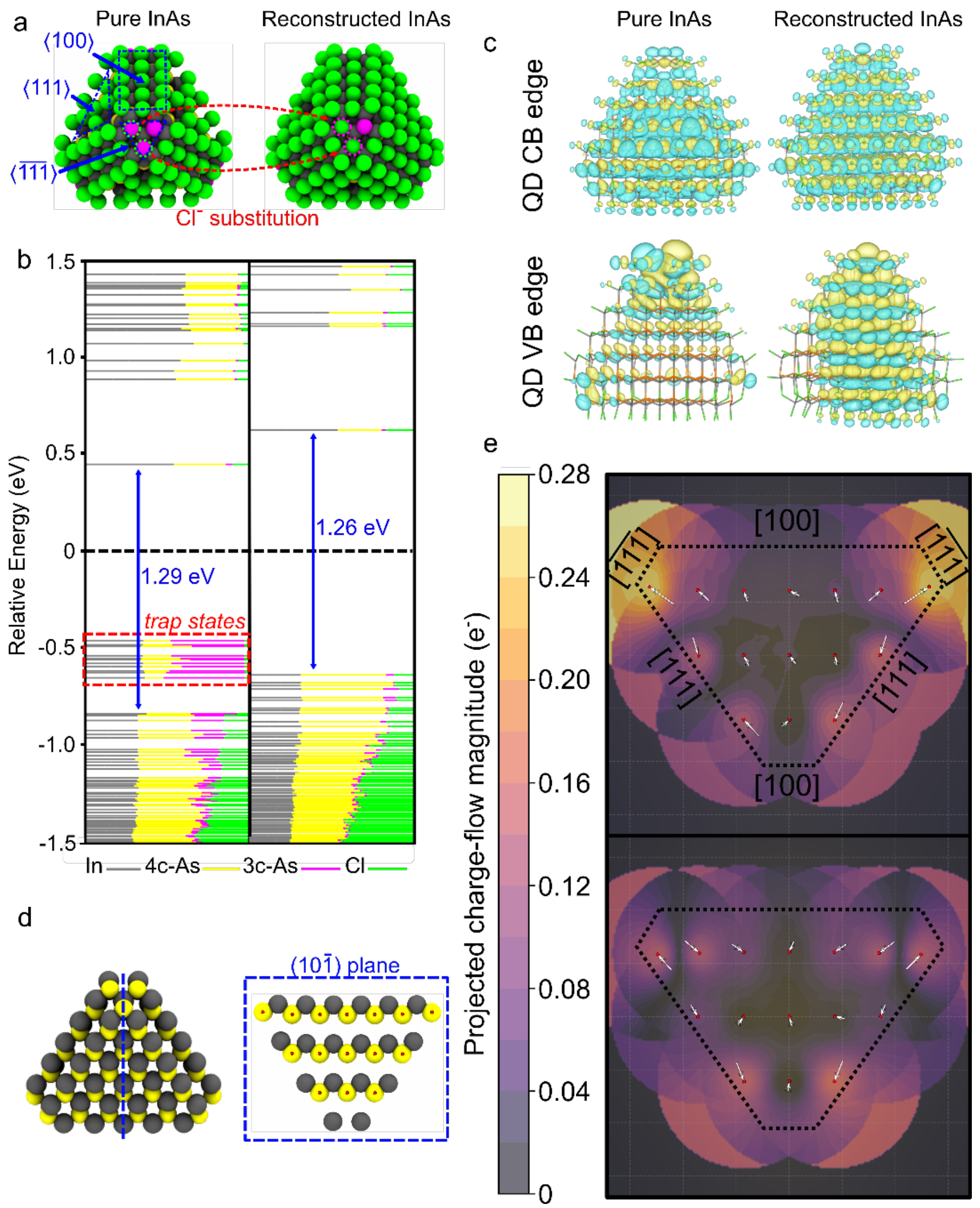

**Figure 1.** Electronic structure and charge accumulation in InAs QD models. (a) Atomic structures of pure (left) and reconstructed (right) 2.1nm InAs cores, highlighting cation-rich ⟨111⟩, anion-rich ⟨$\overline{111}$⟩ and ⟨100⟩ facets, along with an illustration of the $Cl^-$ substitution used to reconstruct the ⟨$\overline{111}$⟩ facets. (b) Projected density of states (PDOS) for the pure and reconstructed models, with the optical HOMO – LUMO gap values indicated in blue. Horizontal segments represent the contribution of each atomic element to a molecular orbital (MO). Energy origin is set to the Fermi level and indicated as a black dashed horizontal line. Color code: In (gray), 4c-As (yellow), 3c-As (magenta) and Cl (green). Trap states are encircled in dashed red lines. (c) QD VB and CB edges

wavefunction isosurfaces for both models. (d) Sketch of a $(10\overline{1})$ plane cut (top left) and atoms lying on this plane (top right). (e) Per-arsenide charge-flow maps on the $(10\overline{1})$ plane for pure (top) and reconstructed (bottom) InAs QD models. White arrows represent the charge-flow vectors: their direction indicates electron flow into As atoms and their length reflects the magnitude. The filled contour background shows the interpolated magnitude on the plane. Dashed lines delimit the QD shape on $(10\overline{1})$ plane. Red dots are a guide to the eye for the projected charge-flow magnitude of As atoms, which are the most sensitive to trap formation. Facet orientations are labeled in the top panel.

To clarify how surface reconstruction improves carrier delocalization, we examined the charge flow along the In-As bonds. To simplify the explanation, we focus solely on As anions (**Figure 1d,e**) so that we can directly elucidate the changes induced by replacing 3c-As with 3c-Cl. Depending on the atomic environment, each ion acts as a central atom that sustain charge flows due to asymmetries in accepting (donating) electrons from (to) neighboring atoms. The asymmetries emerge from being close/far from the surface or near specific facets making almost each As ion truly unique. Analyzing the charge-flow map in **Figure 1e – top** we notice that in the core and subsurface sites, the charge flowing towards As ions is around ~$0.02e^-$ – ~$0.08e^-$, as evidenced by the darker colors. However, it is noticeable a clear direction of charge towards the $\langle\overline{111}\rangle$ facets, which is accumulated by 3-coordinated 3c-As ions with a value of ~$0.27e^-$each. The accumulation is highlighted by the brighter yellow color compared to rest of the QD and translate into a destabilization of the MOs centered on the *4p* of As ions, shifting them within the bandgap (**Figure 1b – left**). Conversely, when reconstructing the facets by replacing As with Cl ions, these facets exhibit generally weaker charge flows, as the introduced

Cl atoms now accept a lower charge, ~$0.13e^{-}$ (almost half of that experienced by 3c-As) (**Figure 1e – bottom**), reflecting the elimination of trap states and delocalizing more effectively the HOMO wavefunction (**Figure 1b,c – right**).

Altogether, the comparison between the pure and surface reconstructed InAs models confirms that surface chemistry strongly controls trap formation and carrier localization on single-material, core-only systems, raising the question of whether such improvements persist once a second material – here, a CdSe (II-VI) shell – is added. At this point, the relevant factor is expected to be not only the surface but also the core-shell interface. To probe this, we focus on charge-balanced InAs@CdSe models with 3ML CdSe, a shell thick enough to prevent too much mixing between the interface and the outer shell surface, while keeping the QD model computationally accessible. In this regime, we can compare two cases on equal footing: an abrupt interface, where the core and shell switch suddenly from one material to the other, and an alloyed (mixed) In-Cd-Se-As interface, yielding a core@interlayer@shell, a configuration already reported for III-V@II-VI architectures[19,25,26,43–45]. In addition, we keep the CdSe outer surface reconstructed in all instances by employing a cation(anion)-vacancy pattern previously developed for II-VI QDs[7], so that any observed differences in the electronic structure can be attributed specifically to the nature of the core-shell interface.

First, we consider the abrupt-interface boundary (**Figure 2a**). In this case, although we reconstructed the outer CdSe shell and eliminated facet-specific traps in that region, the HOMO – LUMO gap still collapses, with both frontier orbitals exhibiting the character of deep trap states driven by a dipole-induced type-III alignment (**Figure 2b, c**). To elucidate the mechanistic origin of this alignment, we analyzed the projected charge-flow magnitudes experienced by the anions of the core@shell model (**Figure 2d**). Here, we can readily observe how the charge is coherently flowing from the core-shell interface to the non-stoichiometric facets in the outer shell, where a large amount of charge ultimately accumulates. This specific surface localization, and not at the interface, arises because the fully coordinated atoms at the interface facilitate charge transfer rather than trapping it; effectively, the interface is a facilitator that drives the excess charge toward the QD boundary, where the lower coordination of surface atoms favors the formation of localized trap states (**Figure 2c**). Consequently, this outflow of charge towards the outer shell surface leaves the InAs core positively charged – shifting its energy levels downward – and the CdSe shell negatively charged – shifting its energy levels upward, as schematized in **Figure 2e**. These anomalous energy shifts in the core and shell account for both the bandgap collapse and the spatial distribution of the HOMO and LUMO wavefunctions. In particular, an expected quasi-type-II band alignment instead becomes a dipole-induced type-III alignment, generally observed across all abrupt-interface, (un)reconstructed-surface models (**Figure S7-S9**), in which the conduction band minimum of the core drops below the valence band maximum of the shell (**Figure 2e**). Although the quantum dot remains globally charge-balanced, this extreme band staggering leads to the population of lower-energy core conduction states at the expense of higher-energy shell valence states. This results in an apparent self p-doping of the shell and concomitant self n-doping of the core, with the Fermi level pinned within the partially empty shell valence band (**Figure 2b**). However, this p-doping is not a real doping effect but rather an artifact arising from the large interfacial dipole. Indeed, this dipole-induced type-III alignment is not observed experimentally[46–48] , which indicates that an abrupt interface between core and shell is highly unlikely; a point highlighting the need to consider modifications at the interface.

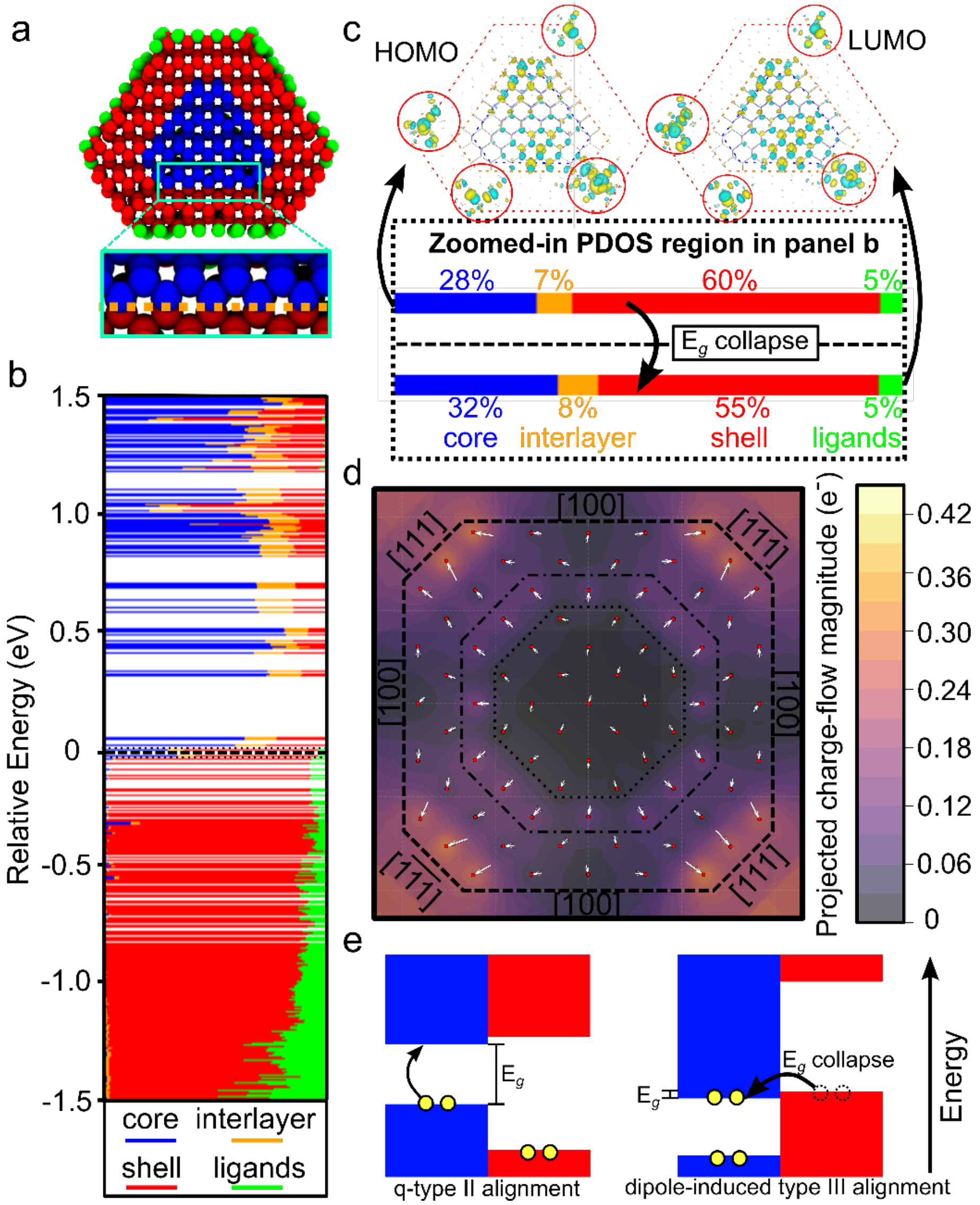

**Figure 2.** Electronic and charge-flow features of the InAs@CdSe QD model with an abrupt core-shell interface and reconstructed CdSe surface (3ML). (a) Atomistic model of the InAs@CdSe QD showing the abrupt interface

between the InAs core (blue) and the CdSe shell (red); green spheres denote Cl ligands. The inset highlights the abrupt boundary between core and shell additionally remarked with a dotted orange line. (b) PDOS decomposed by domains: core (blue), first CdSe ML at the interface (orange), shell (red) and ligands (green). The abrupt interface leads to a collapsed HOMO – LUMO gap. The dotted black area represents the energy window zoomed-in for identifying HOMO and LUMO in panel (c) and the dashed black line points out the Fermi level. (c) Detailed analysis of the frontier molecular orbitals from the zoomed-in PDOS region in (b). Top: HOMO and LUMO isosurfaces divided into core (blue), interlayer (orange) and shell (red) regions by dashed lines, with solid red circles highlighting the wavefunction portions localized at the shell. Bottom: Corresponding composition bars quantifying the percentage of charge density distribution across these domains, including the ligands (green). (d) Per-anions (As and Se) projected charge-flow map on the (001) plane illustrating the crystallographic directions (the corresponding cation map is shown in Figure S5). Arrows indicate donor-to-acceptor directions, while the contour map colors denote the magnitude of the flow. (e) Schematic band-alignment diagram illustrating the effect of the interfacial dipole on the delocalized band edges. The charge redistribution shifts the InAs core levels (blue) to lower energies and the CdSe shell levels (red) to higher energies. This leads to a transition from a quasi-type-II alignment to a dipole-induced type-III alignment where the InAs core conduction band edge overlaps with the CdSe shell valence band edge, resulting in the collapse of the optical bandgap $E_g$. Yellow circles represent electrons, while dotted circles their original positions prior to charge transfer.

To address this, we prepared an alternative core@shell InAs@CdSe system with a reconstructed interface (**Figure 3a-i**). Inspired by the vacancy patterns used for reconstructing QD surfaces in reference[7], we prepared a simple, yet effective interface configuration where all ions of the core-shell, In-As-Cd-Se, are mixed within one monolayer (**Figure 3a-iv**). This configuration approximates the graded interlayer formation observed in InAs@ZnSe heterostructures synthesized at high temperatures[19,32], and parallels the substantial In-diffusion from the InAs core into the CdSe shell confirmed in InAs@CdSe nanowires[49]. In essence, core and shell atoms blend together along with the addition of $\langle 111 \rangle - (2 \times 2)$ cation vacancy patterns[50,51] (**Figure 3a-iii**) showing an overall stoichiometry of $In_{24}As_{31}Cd_{124}Se_{111}$, which roughly approximate to $(InAs)_{0.22}(CdSe)$. This approach helps to reduce the excessive number of Cl ligands that are usually required to maintain charge balance at the outer surface in III-V@II-VI systems (see **Figure S11** in the **SI** for a quantitative ligand density analysis). Once these reconstructions are applied, and the blended interlayer is formed, substantial improvements are observed in the core@shell model. The overall charge-flow magnitudes decrease markedly, and the charge flows appear more randomly distributed, yielding a more homogeneous distribution of charge across the model (**Figure 3b**). As a result: (i) the HOMO – LUMO gap is now present and is of ~0.90 eV (**Figure 3c**); (ii) the LUMO is mostly localized within the InAs core, extending slightly into the shell but not reaching the CdSe surface (**Figure 3d; top**); and (iii) the HOMO remains confined in the InAs core, exhibiting some leakage into the interface region but not into the shell (**Figure 3d; bottom**).

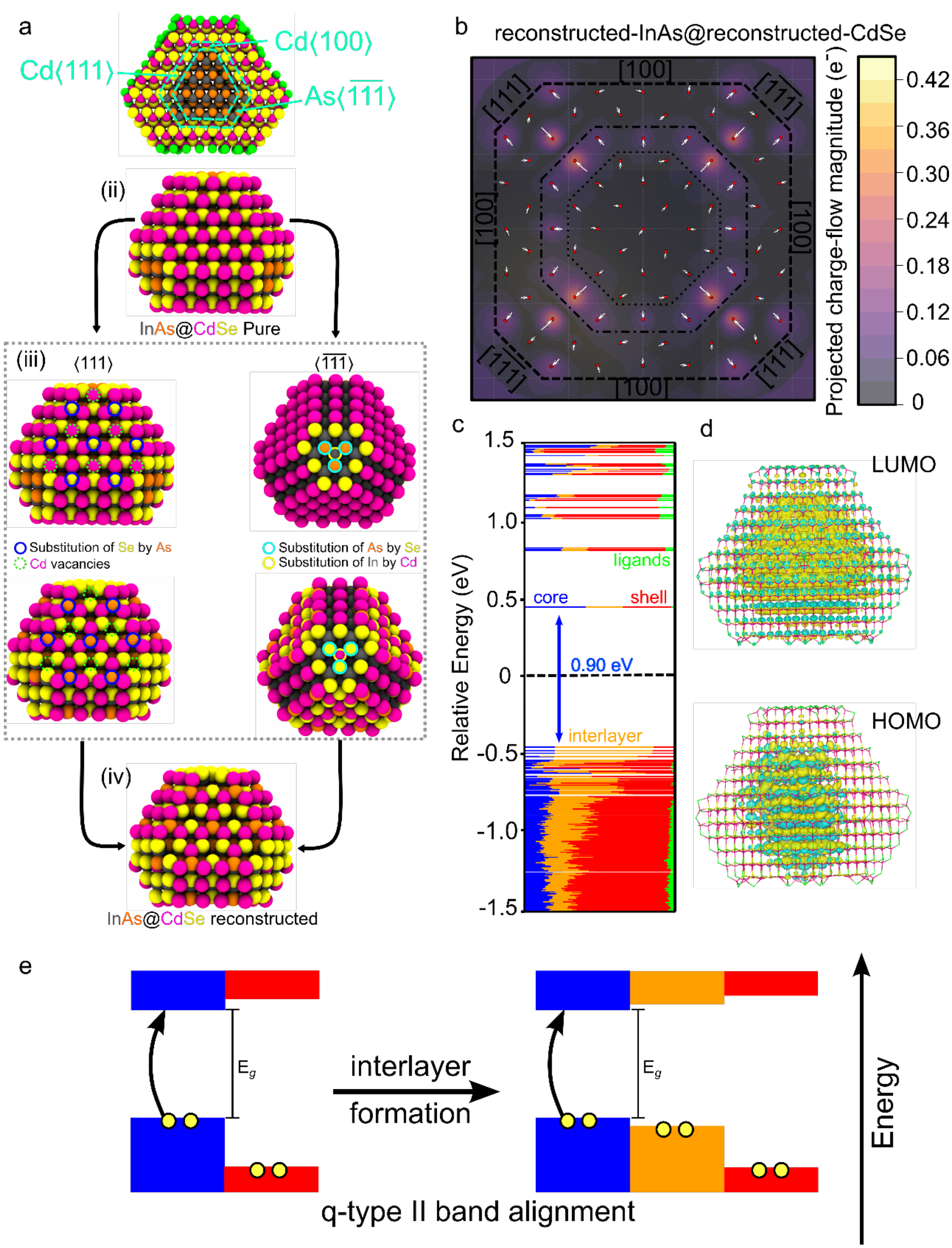

**Figure 3.** Electronic structure of a InAs@CdSe core@shell QD model with interface reconstructions. (a) Atomistic structures illustrating the interface-reconstruction strategy: (i) inner view of a pure core@shell system indicating the interlayer composition, Cd-rich at ⟨111⟩ and As-rich at ⟨$\overline{111}$⟩, and shape (turquoise dashed lines); (ii) atomic

structure of an InAs@(1ML)CdSe model employed for creating the interlayer; (iii) interface reconstruction patterns at: (left) ⟨111⟩ facets combining Se-by-As substitutions (empty dark blue circles) and Cd vacancies (empty green dotted circles), and (right) $\langle\overline{111}\rangle$ facets with As-by-Se (empty cyan circles), and In-by-Cd (empty yellow circles) substitutions; (iv) atomic structure of the resulting InAs@interlayer system. (b) Per-anions projected charge-flow map on the (001) plane illustrating the crystallographic directions. Arrows indicate donor-to-acceptor directions and contour map colors denote charge-flow magnitude. (c) PDOS decomposed by domains: core (blue), blended interlayer (orange), shell (red) and ligands (green). Note that the HOMO states arise primarily from core and interlayer contributions, while the LUMO states delocalize across all domains. The HOMO – LUMO gap value is indicated in blue. The Fermi level is indicated by a black dashed horizontal line. (d) LUMO (top) and HOMO (bottom) wavefunction isosurfaces. (e) Simplified band-alignment diagram illustrating the effects of including an interlayer in the InAs@CdSe QD model based on the domain contributions in panel (c).

Having established that abrupt core-shell interfaces cannot be stabilized by surface reconstructions alone, and that interface reconstruction is essential to recover a trap-free electronic structure, we next evaluate the role of shell thickness under these optimized conditions. Here, we intend to isolate the effect of adding successive CdSe MLs in our models (**Figure 4a**). As shown in **Figure 4b**, the HOMO – LUMO gap remains wide-open across 1-3 ML CdSe shells, decreasing monotonically with shell thickness (**Figure S12**). In all cases, both HOMO and LUMO wavefunctions are delocalized, but their spatial distributions evolve differently with shell growth. For 1 ML, either HOMO or LUMO spreads across the core and shell, although LUMO appears more uniformly distributed throughout the entire system than the HOMO, see **Figure 4b,c (left)**. At 2ML, the LUMO remains mostly confined with the InAs core but extends slightly deeper into the shell, whereas the HOMO is more core-centered and overlaps with the interlayer (**Figure 4b,c – middle**). By 3ML, as noted earlier, this trend continues: the LUMO remains delocalized, occupying mainly the InAs core and spreading slightly to the shell, while the HOMO – similar to the 2ML CdSe shell case – tends to be more pinned at the interlayer (**Figure 4b,c – right**). This evolution in the spatial distribution of the frontier wavefunctions results in a quasi-type-II alignment (**Figure 3e**), consistent with the gradual saturation of the HOMO – LUMO gap, in which the LUMO electrons become progressively less influenced by the shell boundary, while HOMO holes remain confined to the same region regardless of shell thickness[19,48,52]. Comparing these optimized structures with the abrupt models also confirms that internal and external structural reconstruction dominates the electronic properties over surface passivation, as higher ligand densities do not prevent bandgap collapse (**Figure S13**). Finally, while this analysis focuses on the lattice-matched InAs@CdSe system, we hypothesize that the mixed-interface strategy might be equally critical for strained systems, as it is for the case of InAs@ZnSe system, where a mixed In-Zn-Se interlayer spontaneously forms to eliminate interfacial strain; our results suggest this interlayer formation might be driven by a combination of both lattice and electrostatic mismatch. While confirming this interplay requires further targeted investigations, the interlayer engineering we propose likely addresses a broader requirement for stability in III-V@II-VI heterostructures, balancing both chemical dipoles and lattice mismatch.

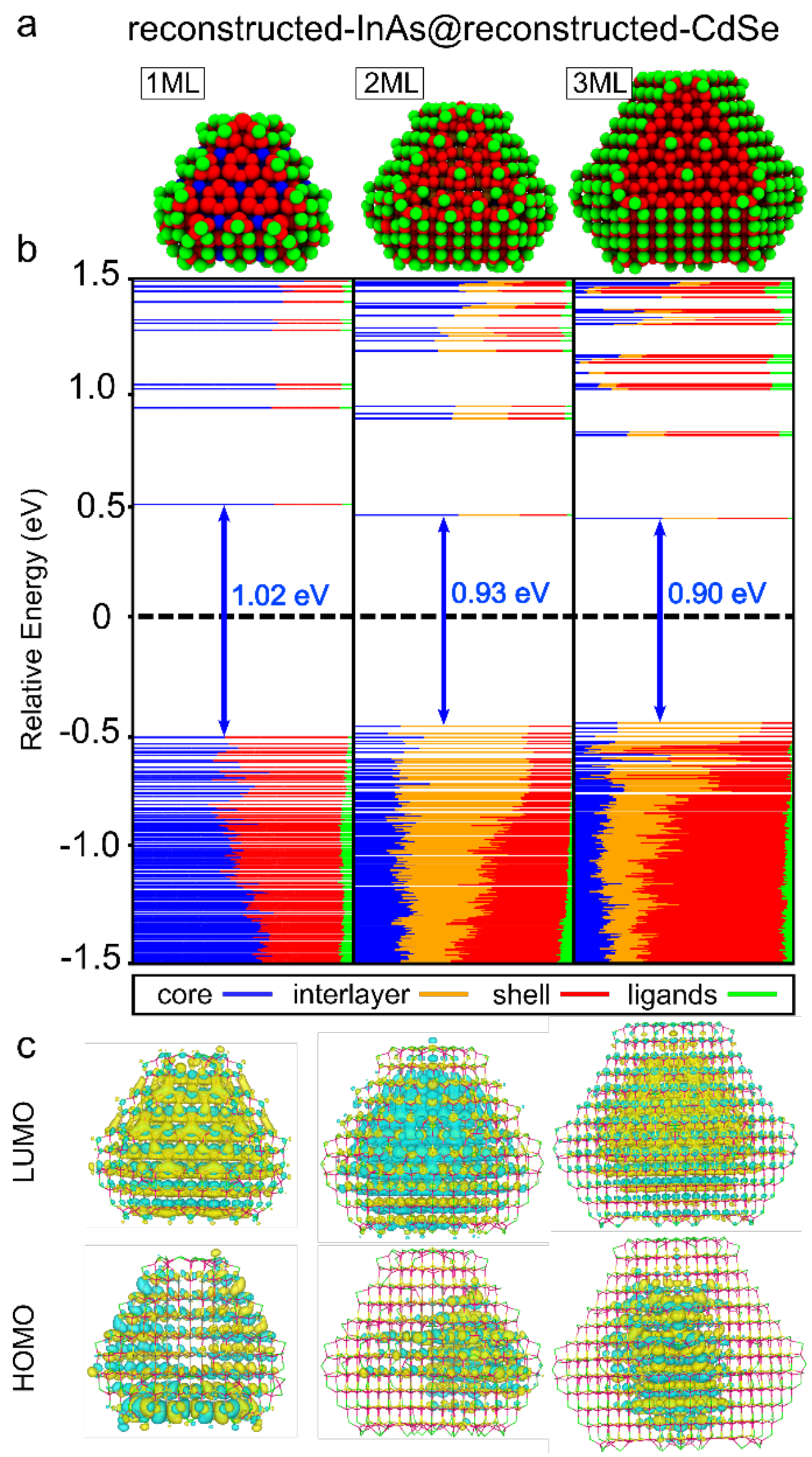

**Figure 4.** Evolution of the electronic structure and frontier orbitals in InAs@CdSe core@shell QD models with increasing shell thickness. (a) InAs@CdSe atomistic models with 1, 2 and 3 CdSe shell MLs. (b) PDOS with color-coded contributions from core (blue), interlayer (orange), shell (red) and ligands (green). The HOMO – LUMO energy gaps are indicated in blue arrows. The Fermi level is indicated by a black dashed horizontal line. (c) Isosurfaces of the LUMO (top row) and HOMO (bottom row) wavefunctions.

In summary, our atomistic models reveal that trap formation in InAs@CdSe QDs cannot be attributed solely to surface coordination but is strongly influenced by the atomic structure of the core-shell interface. Models with only reconstructed outer surfaces still exhibit substantial charge accumulation at the surface, and issue that is mitigated by reconstructing the core-shell interface. The resulting electronic structures display wider gaps with a monotonic dependence on shell thickness, delocalized wavefunctions, and diminished charge-flow magnitudes, yielding more realistic representations of QD heterostructures. Importantly, charge-flow analysis provides direct insight into charge redistributes across the QD, clarifying where traps emerge, and thereby offering a mechanistic foundation for designing improved atomic patterns at either surface or interface. Although the occurrence of reconstructions at the surface and at the interface in III-V@II-VI QDs is now well-established, the detailed atomic structure of the interface remains largely unknown. In this context, computational studies using realistic models of core-shell interfaces, supported by charge-flow analysis, can provide valuable insights on how these patterns affect the overall quality of the heterostructure. While InAs@CdSe serves as an ideal model without interfacial strain, future work on RoHS-compliant systems (e.g., InAs@ZnSe) must also address the interplay between dipoles and interfacial strain. Our results suggest that rational interface engineering could guide new synthetic strategies toward trap-resistant III–V@II–VI QDs with enhanced optoelectronic performance, emphasizing the need for experimental focus on the interlayer region accessible via techniques like elemental analysis or X-ray photoelectron spectroscopy, or solid-state NMR.

## ASSOCIATED CONTENT

The Supporting Information is available free of charge at …

Additional computational details on DFT settings (S1); derivation of the charge-flow vector fields and their connection to classical electric dipoles (S2); supplementary InAs QD models with PDOS and HOMO/LUMO isosurfaces, ligand-density analysis across models, correlation between band-gap values and ligand density along their corresponding charge-flow maps and stoichiometries and elemental ratios (S3); charge-flow maps for cations and anions in pure-pure, pure-reconstructed and reconstructed-reconstructed interlayer-surface InAs@CdSe (S4); additional InAs@CdSe core@shell models (1–3 MLs) with pure-pure interlayer-surface (PDOS, HOMO/LUMO isosurfaces, anion/cation charge-flow maps) (S5); additional InAs@CdSe models (1 and 2 MLs) with pure-reconstructed (PDOS, HOMO/LUMO isosurfaces); additional InAs@CdSe models (1–3 MLs) with reconstructed-pure interface-surface (PDOS, HOMO/LUMO isosurfaces, anion/cation charge-flow maps); stoichiometries and elemental ratios of all the InAs@CdSe models along with their ligand densities; and band-gap evolution versus shell thickness (ML) for all the interface-surface InAs@CdSe core@shell systems (PDF).

A free of charge and open-source code to perform charge-flow analysis from DFT calculations can be found on the GitHub repository: https://github.com/nlesc-nano/charge_flow

## AUTHOR INFORMATION

**Corresponding Author**

jordi.llusar@bcmaterials.net

ivan.infante@bcmaterials.net

**Present Addresses**

†If an author's address is different than the one given in the affiliation line, this information may be included here.

**Notes**

The authors declare no competing financial interest

ACKNOWLEDGEMENTS

J. L., I. I. and A. E. acknowledge Horizon Europe EIC Pathfinder program through the project 101098649 – UNICORN and IKUR Strategy under the collaboration agreement between Ikerbasque Foundation and BCMaterials on behalf of the Department of Education of the Basque Government. We thank the EuroHPC Joint Undertaking for awarding this project access to the EuroHPC supercomputer LEONARDO, hosted by CINECA (Italy) and the LEONARDO consortium through a Regular EuroHPC with number EHPC-REG-2023R03-106. We also thank the Donostia International Physics (DIPC) Supercomputing Center, for which the authors acknowledge for the technical and human support. L. M. and L. D. T. acknowledge the Italian Ministry of University and Research (FIS-2023-03302 IRIDE). Z.H. acknowledges the FWO-Vlaanderen (research project G0C5723N, SBO project S000626N) and Ghent University for research funding (BOF-GOA, 01G02124). Z.H. and I.I. acknowledge research funding from the European Commission (MSCA-DN Track The Twin, grant agreement 101168820)